# Precision Enhancement of Distribution System State Estimation via Tri-Objective Micro Phasor Measurement Unit Deployment

Arya Abdolahi[1], Navid Taghizadegan Kalantari[1]
Department of Electrical Engineering, Azarbaijan Shahid Madani University, Tabriz, Iran

Abstract: A tri-objective optimal Micro Phasor Measurement Units (μ-PMUs) Placement method is presented, with a focus on minimizing the following three parameters: i) the total number of μ-PMU channels, (ii) the maximum state estimation uncertainty, and (iii) the sensitivity of state estimation to line parameter tolerances. The suggested formulation takes single-line and μ-PMU failures into consideration while guaranteeing the complete observability of the system in the presence and absence of contingencies. It also takes into account the impact of zero injection nodes and the quantity of μ-PMU channels carried out at every node. The suggested placement issue is addressed using a customized version of the nondominated sorting genetic algorithm II (NSGA-II). According to the results achieved utilizing three test systems of varying sizes, μ-PMU channels beyond predetermined thresholds only result in higher costs and negligible further decreases in state estimation uncertainty and sensitivity to line parameter tolerances. Additionally, we may omit to instrument between 30 and 40% of buses if μ-PMUs with only two three-phase channels are utilized, with only a modest negative effect on state estimate performance even in the event of contingencies.

## 1. Introduction

μ-PMUs are smart devices used to received the estimation of system phasors. Since the 1990s, μ-PMUs have usually been used in power transmission systems and have been crucial in protecting systems by quickly identifying faults or other imminent critical operational situations. Unfortunately, the M-PMU placement problem is a sensitive problem complicated by the requirement to ensure the system's full observability. This is according to the high device cost and the massive amount of information gathered by the μ-PMU [1].

A single objective function is used in the majority of optimal μ-PMU placement problem formulations, as was mentioned in the literature. Initial attempts to solve the optimal μ-PMU placement problem used standard integer (typically binary) linear programming optimization techniques, as both the total cost of μ-PMU deployment and topological observability are typically linearly proportional to the number of μ-PMUs [2]. Author in [3] proposes an optimal PMU placement problem with the integer linear programming model. In [4], a different strategy was put forth to reduce the quantity of µ-PMUs required to achieve full observability while enhancing state estimation of system by considering the measurement redundancy index. For the purpose of identifying inaccurate data in state estimation, [5] proposed a similar distinction between critical and redundant measurements.

Due to the incorporation of more complex and varied operational circumstances and restrictions, the fundamental optimal µ-PMU placement issue has become more complex over time. The findings and the influence of µ-PMU placement in the presence of various contingencies may be significantly impacted by considering µ-PMU channels number [6]. With these arrangements, the observability constraint must now take into account a greater number of situations [7]. As a result, adding such conditions to the issue formulation causes the connection matrix's row count, which is utilized to apply the observability requirement, to rise quickly. In distribution networks, which typically include a high number of buses, this condition may be particularly challenging to handle. Due to the restricted number of channels, the lines in this study that cannot be directly viewed by a µ-PMU are chosen a priori.

The majority of the factors mentioned above are now included in the optimal µ-PMU placement issue formulation inside a special framework. Furthermore, the goal function, which is provided by a linear or quadratic combination of numerous components, has progressively become more complicated. However, this function frequently ignores the influence of µ-PMU performance entirely and only considers economic and observability factors. Just to provide a few instances, an integer quadratic programming optimal µ-PMU placement model is presented in [8], and it consists of two parts including the level of redundancy and other one is the investment cost of µ-PMUs. By increasing network observability and reducing sensitivity to grid factors, the cost function used in [9] seeks to reduce the µ-PMUs installed in the system. The goal function described in [10] incorporates extra parameters that take into consideration the costs of network unobservability and redundancies to improve system observability in both normal operating settings and contingencies, in addition to the overall deployment costs.

The fundamental drawback of single-objective optimization is the need to combine many values into a scalar cost function in order to find a single solution. The typical method for doing such an aggregate is to balance the goals taken into account by the issue. The results of this method, however, heavily depend on the weights that are selected, which may ultimately result in a lack of variety [11]. One of the major areas for future study in this area is now thought to be "the generalization of the optimal μ-PMU placement issue considering various goals including not just installation cost, but also redundancy, performance, and other design restrictions" [12]. Currently, only a small percentage of optimal μ-PMU placement research studies, the majority of which have two goals, are multi-objective formulation-based. The reduction of the number of μ-PMUs and the maximizing of measurement redundancy—two objectives that are in fact in opposition to one another—are what determine where the μ-PMUs are placed in [13].

A thorough overview of available methods is provided in [14] with regard to the OPP issue solution. A customized edition of the NSGA-II algorithm is implemented in this paper, despite the fact that a number of heuristic algorithms [15] can be used to find satisfactory sub-optimal solutions in a reasonable time horizon. The technique used in earlier works [37], [39] is consistent with the employment of a genetic algorithm for multi-objective OPP. Due to the [16] and [17], NSGA-II is a proper optimization algorithm to μ-PMU placement problem with high accuracy. The ease with which a decent starting population may be generated and the application of penalties to exclude impractical solutions are two further crucial aspects that make the NSGA-II technique especially appropriate in the situation at hand.

Most studies use the single objective function of optimal μ-PMU location in the transmission networks, but this paper presents some contribution as follow:

- Define a multi-objective optimization model with compromising objectives
- Include some limitations for complete observability in the presence of contingencies, kind and number of measurements at every node
- Concentrate on distribution networks that are an emerging field of application for μ-PMUs.

The objective functions taken into account in this study are the total number of μ-PMU channels, the maximum state estimation uncertainty, and the maximum sensitivity of state estimation uncertainty to line parameter tolerance limits. The proposed problem are investigated under the various contingency condition like line outage or M-PMU failure. The μ-PMU sites are discovered using a customized version of the Nondominated Sorting Genetic Algorithm II (NSGA-II), which minimizes the Pareto frontier of the collection of potential solutions, given the tri-objective nonlinear formulation of the current issue [18], [19]. The selected μ-PMU placement technique is suitable for the distribution system because the number of buses and equipment installed in the system is high than in the transmission network.

The remainder of the essay is organized as follows. The tri-objective optimal PMU placement problem is formalized in Section 3, and following a quick summary, each function is represented and supported. The proposed placement strategy's results are reported in Section 4 while taking into account various constraints. Conclusion presented in Section 5.

## 3. Problem formulation

In the case of a grid with $N$ buses and $L$ lines, let $x \in X = \{0,1\}^N$ be a binary vector that $i$th element $x_i$ is adjust to 1 or 0. C(x) shows the available μ-PMU channels number to monitor the network status. The maximum state estimation uncertainty is presented by the U(x). Maximum sensitivity related to the tolerances of network parameter is presented by the S(x). With these explanations, the suggested tri-objective optimal μ-PMU placement problem are formulated as (1).

$$\min_{x \in X}(C(x), U(x), S(x)) \tag{1}$$

There is considered some limitations for the proposed objective function as below.

1) Ensure the complete observability based on the measurement devices and ZINs without considering the effect of contingency condition.

$$A(x + u) \geq 1 \tag{2}$$

In inequality constraint (2), the parameter **A** shows the $N \times N$ undirected graph modeling binary connectivity matrix of the proposed network. As regards, if a μ-PMU failure and branch outage happens, a harder observability limit that (3) is required.

$$\begin{bmatrix} \tilde{A}^1 \\ \vdots \\ \tilde{A}^N \end{bmatrix}(x+u) \geq 1 \quad \text{and} \quad \begin{bmatrix} \tilde{A}^1 \\ \vdots \\ \tilde{A}^L \end{bmatrix}(x+u) \geq 1 \tag{3}$$

where matrix $\tilde{A}^b$ is achieved by substituting the $b$th column of **A** with an all-zero vector.

2) The maximum number of $n_{ci}$ μ-PMU channels that are available and the kind of measurement both affect the μ-PMUs that may be carried out at the node $i$. In many optimal μ-PMU placement articles, the quantity of μ-PMU channels is ignored, and this has to be emphasized. In actuality, not every current phasor is really tracked and utilized for state estimate. This work, on the other hand, takes into account this fundamental technology-related restriction. The restriction on the number of permitted measurements may be modeled as (4).

$$[I_N \quad \Gamma]\begin{bmatrix}x\\x\end{bmatrix} \leq n_c \tag{4}$$

where channels number of measurement devices are shown by $n_c$. $I_N$ is the $N \times N$ identity matrix, and the measurement installed at each node are shown by a binary matrix as $\Gamma$.

*A. First objective function: Minimum number of µ-PMU channels*

The optimal µ-PMU placement is considered for make the µ-PMU placement strategy as generic as feasible. In fact, this quantity not only increases with the number of µ-PMU, but it also significantly affects equipment cost.

$$C(x) = c^T.x \tag{5}$$

where $c \leq n_c$ is the column vector including the number of µ-PMU channels available for both voltage and current measurements at all buses.

*B. Second objective function: minimize the state estimation uncertainty*

The uncertainty related to system state estimation is modeled in the second objective function that relies on the parameter chosen and the desired state estimation. The weighted least squares estimator method considered depends on two hypotheses.
1) The data collected from µ-PMU based on zero injection nodes are utilized for state estimation;
2) As given in [20], the state variables and measurement devices are converted from polar to rectangular coordinates.
The observability of system is only determined by µ-PMUs and zero injection nodes. This hypothesis is completely expected in research papers on optimal PMU placement procedures. The choice to use rectangular variables instead of simplifying the linear system equations used for state estimation comes with associated advantages from a computational viewpoint [21].

Suppose x is a binary vector representing the nodes where µ-PMUs are located. The distribution system state estimation depends on current phasor measurements ($M_I$), voltage phasor measurement ($M_V$), and zero injection node current measurement ($M_Z$). The sum of measurements data ($M = M_V + M_I + M_Z \geq N$) divided into real and imaginary sections presented by $R$ and $I$, and converted to a single 2$M$-long vector ($z = \begin{bmatrix} z_{V_R}^T, z_{V_I}^T, z_{I_R}^T, z_{I_I}^T, 0^T, 0^T \end{bmatrix}^T$). The state variables and measurement data can be presented by the Eq. (6), if the variables of the system state are described as rectangular coordinates and are converted to a 2$N$-long vector ($v = \begin{bmatrix} v_R^T, v_I^T \end{bmatrix}^T$) [20].

$$z = \begin{bmatrix} z_{V_R} \\ z_{V_I} \\ z_{I_R} \\ z_{I_I} \\ 0 \\ 0 \end{bmatrix} = H(x) \begin{bmatrix} v_R^T \\ v_I^T \end{bmatrix} + \varepsilon \tag{6}$$

where $\varepsilon = \begin{bmatrix} \varepsilon_{V_R}^T, \varepsilon_{V_I}^T, \varepsilon_{I_R}^T, \varepsilon_{I_I}^T, \varepsilon_z^T, \varepsilon_z^T \end{bmatrix}^T$ is the vector presenting the random uncertainty contributions effected by various measurements, and

$$H(x) = \begin{bmatrix} \tilde{I}(x) & 0 \\ 0 & \tilde{I}(x) \\ G(x) & -B(x) \\ B(x) & G(x) \\ G_z & -B_z \\ B_z & G_z \end{bmatrix} \tag{7}$$

$$\hat{v} = F(x)z = \begin{bmatrix} H^T(x) R^{-1} H(x) \end{bmatrix}^{-1} H^T(x) R^{-1} z \tag{8}$$

here $\hat{\cdot}$ represents an expected amount. R is diagonal if all components to measurement uncertainty are considered to be uncorrelated. The items related to ZI actual measurements ought to be empty. In order to make matrix R invertible, these values must be swapped out with very tiny amounts that are at least two orders of magnitude smaller than the other nonzero components of R.

Let $e_v^c = (\hat{v}_R - v_R) + j(\hat{v}_I - v_I)$ represent the $N \times 1$ dimensional random complex error vector that is the consequence of zero-mean estimate errors for the real and imaginary sections of the state variables. The following equation illustrates a potential scalar and conservative function for describing the uncertainty of state estimate.

$$U(x) = \max\left\{\sqrt{\mathrm{Eig}(\Phi_v^c)}\right\} \tag{9}$$

where the eigenvalues of the argument matrix are returned by the function Eig(.) [22].

*C. Third objective function: Minimize sensitivity to line parameter tolerances*

It is impossible to describe the sensitivity of system state estimate to the uncertainty influencing grid parameters using a single formulation. For a certain measurement setup, the sensitivity function S(x) in this study is defined as the greatest increase of the members of the covariance matrix of state estimation errors caused by unidentified tolerances of line parameters.

$$\delta H(x) = \begin{pmatrix} 0 & 0 \\ 0 & 0 \\ \delta G(x) & -\delta B(x) \\ \delta B(x) & \delta G(x) \\ \delta G_z & -\delta B_z \\ \delta B_z & \delta G_z \end{pmatrix} \tag{10}$$

The covariance matrix of $e_v$ represented as [23], assuming that the estimate errors of the real and imaginary sections of the selected state variables provided by (8) are rearranged into the $2N \times 1$ vector $e_v = [(\hat{v}_R - v_R)^T, (\hat{v}_I - v_I)^T]^T$.

$$\tilde{\Phi}_v = E(e_v e_v^T) = \tilde{F}(x)E\{\varepsilon\varepsilon^T\}\tilde{F}^T(x) = \tilde{F}(x)R\tilde{F}^T(x) \tag{11}$$

where $\tilde{F}(x) = \left[\tilde{H}^T(x)R^{-1}\tilde{H}(x)\right]^{-1}\tilde{H}^T(x)R^{-1}$. It is necessary to separate the effects of tolerances and measurement uncertainty in (11) in order to assess how line parameter tolerances affect $\tilde{\Phi}_v$ independently of the accuracy of the measurements that are available. If $\sigma_r$ denotes the relative standard uncertainty comment to all measurements, then matrix R cam be written as $R = \sigma_r^2 \tilde{R}$. Therefore, recalling that $\tilde{R}^{-1} = \tilde{R}^{-1^T}$, after a few steps (11) can be written as

$$\tilde{\Phi}_v = \tilde{S}(x)\sigma_r^2 = [\tilde{H}^T(x)\tilde{R}^{-1}\tilde{H}(x)]^{-1}\sigma_r^2 \tag{12}$$

where $\tilde{S}(x)$ can be thought of as the sensitivity matrix since its components show the rate of changes of the entries of the state estimation errors covariance matrix caused simply by the tolerance values.

If the components of (10) are supposed to be equally distributed around the respective nominal values within a specified relative proportion $\pm\Delta$ of $H_{ij}$ for $j = 1,\ldots,2M$ and $i = 1,\ldots,2N$, Eq. (13) shows the maximum sensitivity to line parameter tolerance.

$$S(x) = \max_{i,j=1,\cdots,N} \left\{ \max_{\delta H_{ij} \in [-\Delta.H_{ij}, \Delta.H_{ij}]} \left\{ \tilde{S}_{ij} \right\} \right\} \tag{13}$$

## 4. Simulation and numerical results

To analysis the efficiency of the suggested tri-objective model, some simulation results associated with optimal μ-PMU allocation is given in this section. Three different distribution networks are considered to evaluated the proposed model on them as follow.

- 37-bus distribution network
- 85-bus distribution network [24]
- a portion of Caracas metropolitan area [25] (141-bus distribution network)

Two different case studies are taken into account for the optimal μ-PMU placement problem (shortly called *Case A* and *Case B*).

- *Case A*: In this case $\Gamma = I_N$ and $n_c = 2$. Ii is supposed to a μ-PMU has maximum two three-phase inputs channels. These channels applied to measure the no zero injection current and voltage phasors.
- *Case B*: In this case $\Gamma = A$ and $n_c = 1+2$, which l is the column vector containing the lines number connected to every bus.

Two additional subcases are examined relying on whether contingencies are considered or not. If not any contingencies are taken into account the optimal μ-PMU allocation problem depends only on limitations (2) and (4). Otherwise, limitation (3) replaces (2). In all the case studies, the parameters of the NSGA-II algorithm is used from the parameters suggested in [26]. The probabilities value of crossover and mutation are adjusted to 100% and 10%, respectively, whereas with these amounts, the probability that the optimization algorithm gets stuck in possible local minima is very low. In addition, the amount of generation numbers and the size of the population is selected after several iterations to guarantee a suitable convergence of the Pareto curve.

The generation number hypervolume curves obtained from the optimal μ-PMU location problem presented in Fig. 1 for the three distribution test systems, supposing to utilize 2-channel μ-PMU just without considering contingency (*Case A*). The simulation results for *Case B* are identical to

*Case A* with considering the contingency condition. Simulations were implemented on a Windows 10 laptop configured with Intel(R) Core (TM), CPU i7-8565U, 2.8 GHz, and 16 GB of RAM.

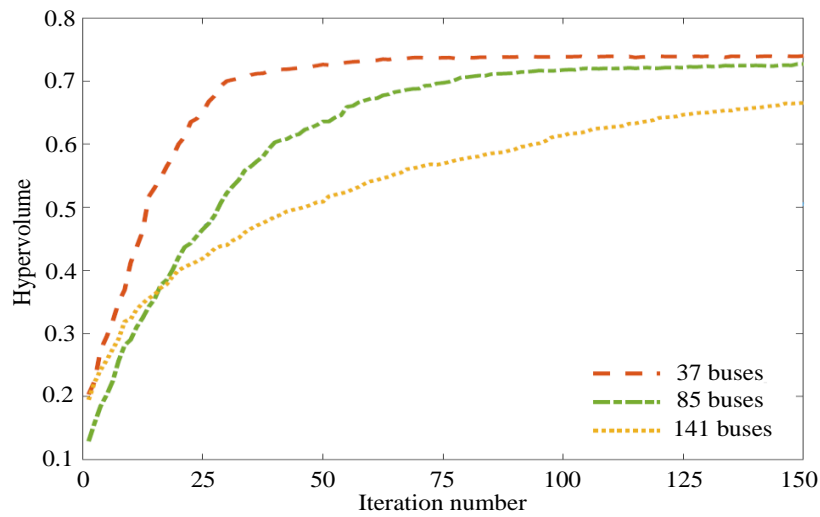

Fig. 1. Hypervolume curves related to the optimal μ-PMU placement

The 3-D Pareto frontiers solutions determined by the NSGA-II are presented in Figs. 2-4 for the three suggested distribution networks, supposing that no limitations of the contingency condition have been taken into account. Every axis of these figures show one of the objective functions (2-4). Nevertheless, the amount of uncertainties are described as a percentage of the nominal slack bus voltage to enhance readability. The Pareto frontiers achieved by containing the limitation of contingency condition are not displayed for the briefness sake, since they are nearly containing in those displayed in Figs. 2-4, as it will be presented briefly. The results supply a qualitative rather than a quantitative outline of optimal μ-PMU location results, It is noteworthy that

- In *Case A*, the restriction on the maximum number of channels and the permitted measurement types clearly limits the number of μ-PMU channels. However, as compared to *Case B*, the state estimate uncertainty and sensitivity do not seem to significantly increase.
- Additional findings, achieved with various amounts of A, demonstrate that the Pareto frontier's maximum state uncertainty amounts scale as anticipated.

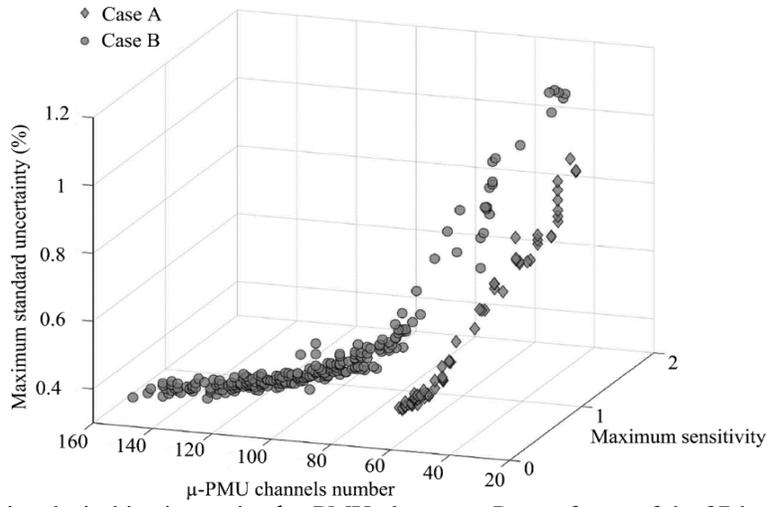
Fig. 2. Three-dimensional tri-objective optimal μ-PMU placement Pareto fronts of the 37-bus distribution system

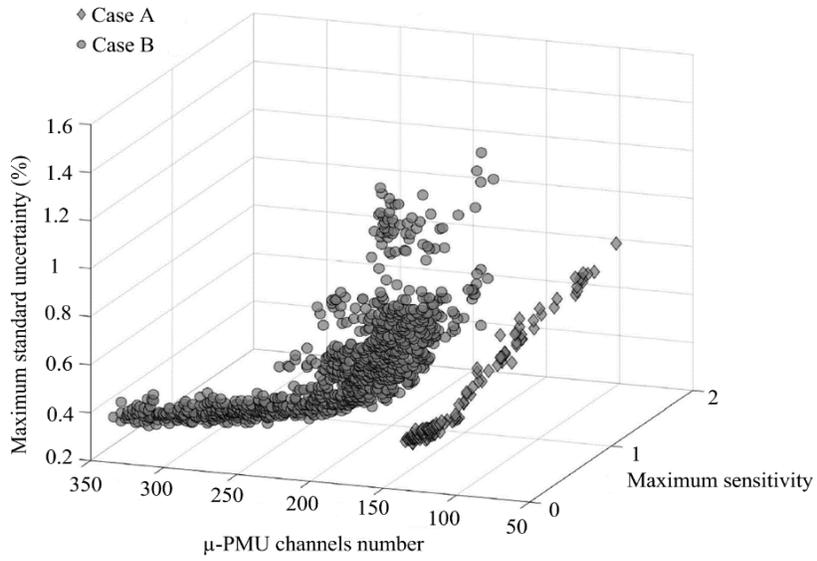
Fig. 3. Three-dimensional tri-objective optimal μ-PMU placement Pareto fronts of the 85-bus distribution system

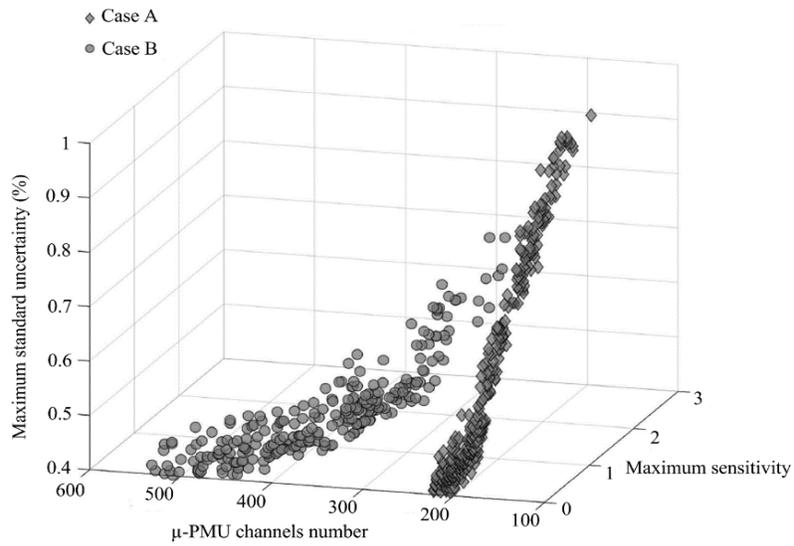
Fig. 3. Three-dimensional tri-objective optimal μ-PMU placement Pareto fronts of the 141-bus distribution system

The trade-offs between the optimal μ-PMU location answers achieved by the suggested tri-objective method can be inferred from the curves displayed in Figs. 4 and 5. The projection minimum envelopes of the 3-D Pareto frontiers onto orthogonal planes are drawn in Figs. 4(a)-(b) and 5(a)-(b), respectively. The results with contingency are shown with dashes lines and the results without considering contingency are shown with solid lines. Some interesting information about the optimal μ-PMU locations are provided by figures 4 and 5 that are summarized as follow.

If no limitations for contingencies are considered, once the number of μ-PMUs is adequate to guarantee the system full observability, the maximum standard estimation uncertainty and the maximum sensitivity tend to decrease very dramatically as the number of μ-PMUs channels increases. Distribution networks with small-scale produce slightly steeper curves. This tendency is due to the fact that state estimation in small systems is more significantly impacted by new μ-PMUs than it is in big systems.

It is crucial to emphasize that the quantity of μ-PMU channels should not be confused with the quantity of μ-PMU locations and, accordingly, with the quantity of measurement devices that need to be placed. As predicted, the curves in Figs. 4(a)-(b) and 5(a)-(b) reveal that *Case B* always has more μ-PMU channels than *Case A*. However, in *Case B*, the minimum number of grid buses being observed by a μ-PMU is typically a little lower than in *Case A*. Particularly, if no limitations for contingencies are included in the issue, system observability can be accomplished by instrumenting 39% ± 5% of network buses in *Case A* and *Case B*, respectively. In addition to network structure, these values also depend on the quantity and arrangement of ZINs within the system.

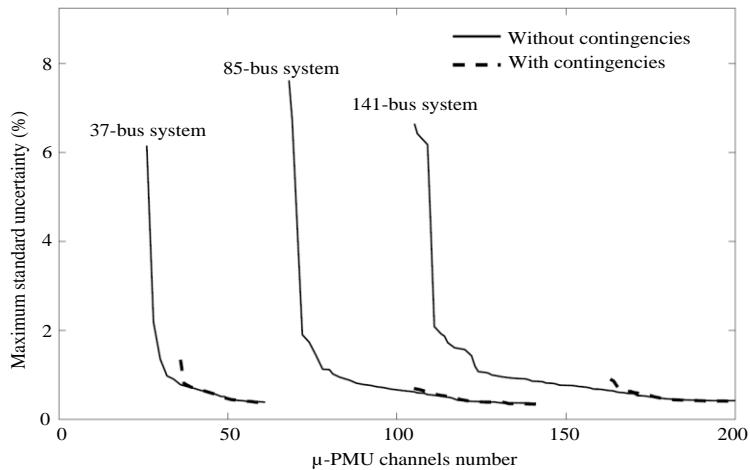

(a)

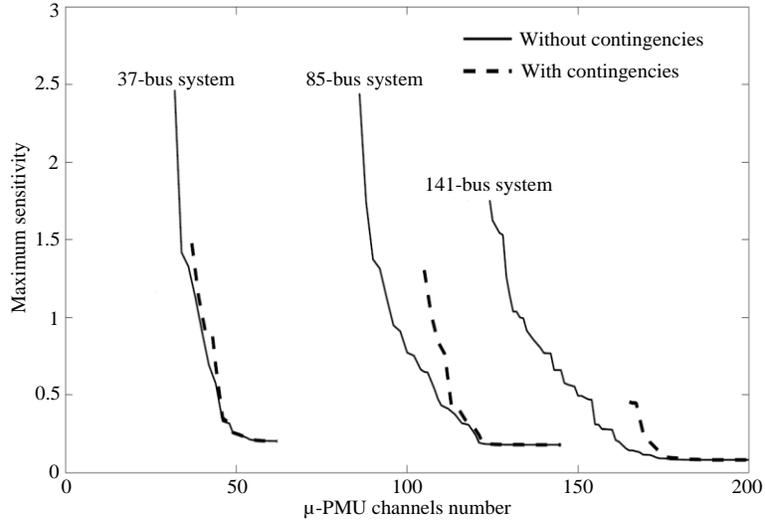

(b)

Fig. 4. Pareto frontiers of *Case A* for optimal PMU placement for three distribution networks

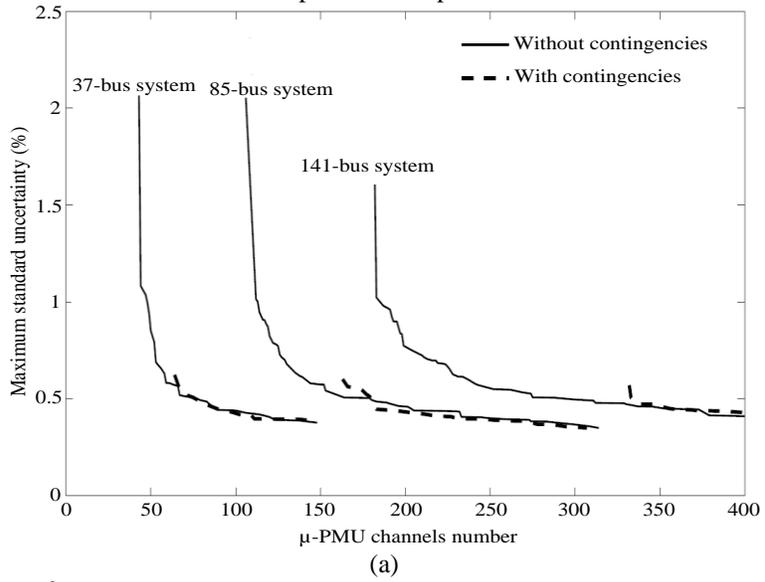

(a)

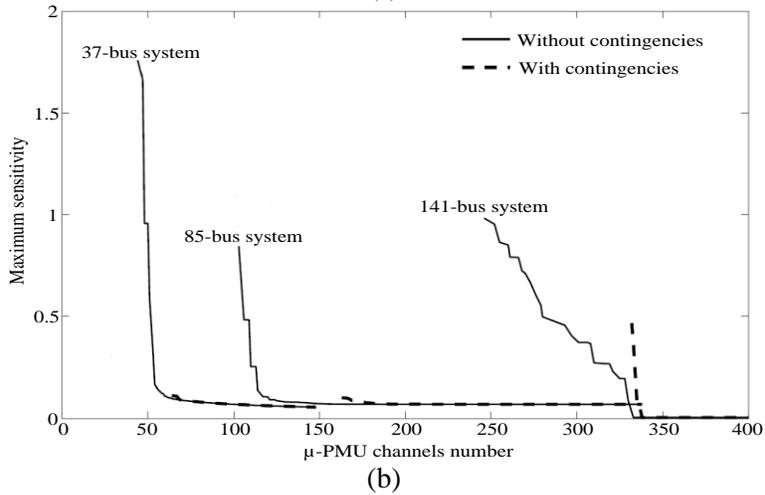

(b)

Fig. 5. Fig. 4. Pareto frontiers of *Case A* for optimal PMU placement for three distribution networks

Optimal μ-PMU placement problem results for three test system are presented as Table 1-3. All of the results are implemented with and without considering M-PMU channels are depicted. As can be seen from these tables, number of μ-PMU channels had have a significant effect on the bus number equipped to the measurement devices. Finally, it is dedicated that the measurement devices with high μ-PMU channels cause to use lees number of measurement devices.

Table 1. Optimal μ-PMU placement problem results for the 37-bus distribution network under two case study

|  | Type of bus | No. of μ-PMU | Bus Numbers |
|---|---|---|---|
| Case A | Without μ-PMUs | 12 | 22, 27, 28, 29, 30, 31, 32, 33, 34, 35, 36, 37 |
|  | With 2 μ-PMUs channels | 25 | 1, 2, 3, 4, 5, 6, 7, 8, 9, 10, 11, 12, 13, 14, 15, 16, 17, 18, 19, 20, 21, 23, 24, 25, 26 |
| Case B | Without μ-PMUs | 17 | 15, 17, 18, 20, 22, 26, 27, 28, 29, 30, 31, 32, 33, 34, 35, 36, 37 |
|  | With 3 μ-PMUs channels | 10 | 1, 3, 6, 8, 9, 12, 16, 19, 23, 25 |
|  | With 4 μ-PMUs channels | 9 | 2, 4, 5, 10, 11, 13, 14, 21, 24 |
|  | With 5 μ-PMUs channels | 1 | 7 |

Table 2. Optimal μ-PMU placement problem results for the 85-bus distribution network under two case study

|  | Type of bus | No. of μ-PMU | Bus Numbers |
|---|---|---|---|
| Case A | Without μ-PMUs | 24 | 2, 3, 5, 7, 8, 9, 10, 12, 13, 27, 29, 32, 34, 35, 41, 48, 49, 52, 58, 64, 65, 67, 70, 81 |
|  | With 1 μ-PMUs channels | 2 | 68, 73 |
|  | With 2 μ-PMUs channels | 59 | All the others |
| Case B | Without μ-PMUs | 38 | All the others |
|  | With 3 μ-PMUs channels | 21 | 1, 15, 16, 17, 36, 38, 47, 51, 54, 55, 56, 59, 62, 66, 72, 74, 75, 78, 82, 84, 85 |
|  | With 4 μ-PMUs channels | 14 | 4, 6, 11, 21, 31, 44, 46, 50, 53, 57, 61, 63, 80, 82 |
|  | With 5 μ-PMUs channels | 11 | 2, 3, 5, 8, 19, 26, 29, 32, 41, 48, 70 |
|  | With 6 μ-PMUs channels | 1 | 67 |

Table 3. Optimal μ-PMU placement problem results for the 141-bus distribution network under two case study

|  | Type of bus | No. of μ-PMU | Bus Numbers |
|---|---|---|---|
| Case A | Without μ-PMUs | 60 | 2, 3, 4, 5, 6, 7, 10, 11, 14, 15, 16, 17, 19, 22, 24, 25, 26, 28, 31, 33, 38, 40, 42, 43, 45, 46, 47, 50, 54, 55, 57, 60, 61, 63, 64, 70, 78, 79, 81, 85, 87, 88, 90, 91, 92, 93, 95, 97, 99, 101, 103, 108, 118, 119, 120, 121, 122, 124, 125, 126 |
|  | With 1 μ-PMUs channels | 7 | 18, 30, 102, 104, 114, 115, 131 |
|  | With 2 μ-PMUs channels | 74 | All the others |
| Case B | Without μ-PMUs | 50 | All the others |
|  | With 3 μ-PMUs channels | 44 | 1, 32, 34, 35, 36, 52, 53, 59, 62, 68, 69, 71, 72, 74, 75, 77, 80, 82, 83, 84, 87, 98, 100, 105, 106, 107, 109, 110, 111, 112, 113, 116, 117, 130, 132, 133, 134, 135, 136, 137, 138, 139, 140, 141 |
|  | With 4 μ-PMUs channels | 36 | 8, 9, 12, 20, 21, 27, 29, 37, 39, 40, 41, 48, 51, 54, 56, 57, 58, 61, 64, 65, 66, 67, 73, 76, 86, 88, 96, 101, 103, 108, 115, 123, 124, 127, 128, 129 |
|  | With 5 μ-PMUs channels | 10 | 13,18,23, 44, 49, 63, 79, 89, 104, 119 |
|  | With 6 μ-PMUs channels | 1 | 94 |

## 5. Conclusion

In this article, the problem of installing phasor measurement units (PMUs) is approached from a novel angle, i.e. by solving a tri-objective optimization problem with the following objective functions: the total number of µ-PMU channels, the maximum state estimation uncertainty based only on high-rate µ-PMU measurements, and the maximum state estimation sensitivity to line parameter tolerances. The issue formulation also takes into account limits on the number of µ-PMU channels, the kind of allowable µ-PMU measurements at each bus, and the system observability (both with and without taking constraints for contingencies into account). A customized version of the genetic algorithm NSGA-II is used to solve the tri-objective optimization issue. Perhaps other heuristic optimization techniques might do even better in terms of computing. The suggested NSGA-II method, however, clearly converges to the Pareto borders of interest in each of the distribution systems under test, therefore no significant changes are anticipated in either the findings or the conclusions.

Since extensive µ-PMU deployments to enable smart grid operation is becoming more popular, our study's emphasis is on distribution systems. Outcomes from three test distribution networks with 37, 85, and 141 buses each show that as the number of buses equipped with µ-PMUs approaches definite tolerances, there is little to no reduction in the maximum state estimation uncertainty and maximum sensitivity to line parameter tolerances, while there is a significant increase in the cost of the instruments. According to the findings from the report's three distribution networks, also with µ-PMUs equipped with just two measurement channels, between 30 and 40 percent of all buses might potentially go unmonitored. If there are no restrictions on how many µ-PMU channels are available, this proportion might increase by a certain percentage. The second option does not, however, often provide significant state estimate improvements and is not typically economically viable. It's rather intriguing to note that an ideal µ-PMU placement configuration capable of ensuring low state estimation uncertainty and sensitivity is likely to be resilient to contingencies as well.